\documentclass[11pt]{article}
\usepackage{amsfonts}
\usepackage{epsfig}
\newcommand{\be}{\begin{equation}}
\newcommand{\ee}{\end{equation}}
\newcommand{\bea}{\begin{eqnarray}} 
\newcommand{\eea}{\end{eqnarray}}

\DeclareMathSymbol{\mg}{\mathrel}{symbols}{"1D}

\newcommand{\cM}{{\cal M}}

\newcommand{\cO}{{\cal O}}

\newcommand{\cR}{{\cal R}}

\newcounter{oldcounter}

\if@twoside\oddsidemargin25mm
\else
\begin{document}
\begin{flushright} 
DAMTP-2002-141.
\end{flushright} 
\vskip3.4cm

\begin{center}
{\Large{\bf{TeV-scale Z$^\prime$ bosons in intersecting  D-brane SM-like models}}}
\end{center}
\vspace{0.83cm} 
\begin{center}
{\bf D.M. Ghilencea}\footnote{This contribution is based on work done in 
collaboration with L.E. Ib\'a\~nez, F. Quevedo and N. Irges.}
\end{center}

\begin{center}
{\it DAMTP, CMS, University of Cambridge \\
Wilberforce Road, Cambridge, CB3 0WA, United Kingdom.}
\end{center}
\bigskip 

\vspace{2.cm}
\begin{center}
{\bf Abstract}\\
\vspace{0.6cm}
 \end{center}
{\small Recent string constructions with intersecting D6 and D5 brane models
succeeded in predicting in the low energy limit a symmetry group
and a fermionic spectrum similar  to that of the SM. In such
constructions additional $U(1)$ 
fields are a generic presence and they (or linear combinations
thereof) become massive through a  string mechanism which couples them 
to 4D RR two-forms. $U(1)_Y$ of hypercharge emerges as a linear
combination of the initial $U(1)$ symmetries while the massive 
$U(1)$ fields induce (through mixing with Z boson) corrections 
on Z mass. $\rho$ parameter constraints on the latter allow one to set
lower bounds on the string scale in the TeV region. These  
can then be used to  {\it predict} lower bounds on the masses of the 
additional $Z'$ bosons, without specific assumptions about the
compactification volume.}

\vspace{0.4cm}
\noindent
{\sl
Contribution to the Proceedings of the \\  
1$^{st}$ International Conference on 
String Phenomenology, Oxford (U.K.), 6-11 July 2002.
}

\newpage\setcounter{page}{1}

\section{Introduction.}

The possibility of (large) extra space dimensions has attracted much
research interest in the context of both string and  effective field 
theories. 
In general the latter can  be successfully  used for 
many theoretical/phenomenological studies in this
direction. However, it is generally thought  that a more complete and 
consistent picture of the  high energy physics is that brought by string theory.  
It is thus desirable to obtain viable field theory models as  direct 
constructions of string theory. For this purpose, recent chiral 
D-brane string constructions with branes located at singularities 
\cite{Quevedo1,aiqu,cuw} or models with D-branes intersecting at 
non-trivial angles
\cite{imr,bgkl,Cremades:2002dh}
have brought in a new approach, alternative to that of (earlier) heterotic
string compactifications. Finding out which of these 
approaches is more successful  in obtaining in the low energy limit
models close to the Standard Model (SM) or its minimal  supersymmetric
extension  (MSSM) is  a rather difficult problem. This is
due to the large variety of different string vacua.  For this reason, 
identifying some generic ``predictions'' of string  compactifications 
may provide  an useful approach to identifying viable string models.

In the following  some phenomenological implications of 
the D-brane models are addressed. To obtain {\it chiral} D-brane 
models one uses 
either models with D-brane at singularities or with 
intersecting branes. These share some common properties: a low value 
for the string scale is allowed, the gauge symmetry contains direct products 
of groups $U(N_\alpha)\times U(N_\beta)$ with $U(N_\alpha)$ to arise
from the stack ``$\alpha$'' of $N$ individual $U(1)$ branes; fermions 
can transform as bi-fundamental  representations  of these groups; 
$U(1)$ group factors in addition to the non-Abelian SM gauge group are
a generic presence and the $U(1)$ of hypercharge emerges as a linear 
combination of these. Other characteristics  are more
model dependent, for example the models may be supersymmetric  or
non-supersymmetric \cite{imr}. The latter  requires fixing  
a low value for the string scale, to avoid re-introducing the 
hierarchy problem.

D-branes models with a spectrum  somewhat close to that of the  SM
include constructions with D4 branes and D5 or D6 branes 
intersecting at non-trivial angles. In the following we address 
the specific class of models with intersecting D6-branes of
\cite{imr}, although some of the results are more general.
The   models  are constructions with D6 branes wrapping 
a 3-cycle on a six torus (orientifolded) type II-A string theory.
It involves  four stacks $\alpha=a,b,c,d$ of D6-branes with gauge
groups $U(3)_a$, $U(2)_b$, $U(1)_c$ and $U(1)_d$, containing
inside the Minkowski space and wrapping each of the remaining three 
dimensions of the branes on a different torus $T^2$.  We denote by  
$n_{\alpha i}, (m_{\alpha i})$, i=1,2,3, $\alpha=a,b,c,d$ 
the  number of times each brane  $D6_\alpha $  is 
wrapping around the  x(y)-coordinate of the $i-th$ torus. This 
construction with four stacks is the minimal one which can accommodate 
a fermionic spectrum similar  to that of the SM case (plus three
generations of  right-chiral neutrinos).  However, the gauge 
symmetry  contains in addition 
to the non-Abelian part of the SM gauge group four additional
$U(1)_\alpha$, $\alpha=a,b,c,d$  which emerge  from the four
stacks of  branes considered.  Such $U(1)$'s  are a generic presence 
in models with intersecting branes and they play a central 
role in the discussion below. (Similar considerations apply to D5 brane
models \cite{Cremades:2002dh} which are type II-B compactifications
on an orbifold $T^2\times T^2\times (T^2/Z_N)$).

An important feature of the models considered is the presence in the
4D action of the couplings $ c_i^\alpha B_i\wedge F_\alpha$, where 
$F_\alpha$ is the field strength of a $U(1)_\alpha$ 
field and $B_i$ is a 4D RR two-form. The  couplings emerge, in the
case of D6 branes from  $C_5 \wedge F$ integrated over a 
three-cycle of the six-torus. Such couplings together with the kinetic 
term for $F_\alpha$ and $B_i$ can be re-written as a kinetic term for 
a $U(1)$ field together with an associated  mass term. This is
possible because the scalar dual to the
$B_i$ two-form is ``eaten'' by the  $U(1)_\alpha$  field which thus becomes 
massive  \cite{Ghilencea:2002da}. The procedure does not require a
Higgs mechanism or  a Higgs field be present in
the spectrum. Note that the  mechanism requires the coupling 
$B_i \wedge F_\alpha$, but not $\eta F_\alpha \wedge F_\alpha$
coupling. For a $U(1)$ field with a Green-Schwarz 
coupling $B\wedge F$,  being anomalous (in four dimensions) is not a 
necessary condition to acquire a mass, and anomaly free $U(1)$'s can
become massive. In previous (heterotic) string models 
 anomalous $U(1)$  also had a Green-Schwarz term and only these 
became massive. Further, since there is no Higgs mechanism, the 
initial (or combinations thereof)
 $U(1)_\alpha$ may survive as a perturbatively exact 
global  symmetry of the models.

This observation can  be useful for phenomenological
purposes. For example the presence of  $U(1)_a$ as a global symmetry 
ensures that baryon number is conserved, thus avoiding proton decay,
a generic problem  in models with a low string/UV cut-off scale. For 
this class of models (diagonal) lepton number  is also an exact
symmetry and Majorana neutrino masses are forbidden.
Dirac neutrino masses are allowed and neutrino oscillations can 
take place (only the sum $L_e+L_\mu+L_\tau$ is an exact symmetry).

An important aspect of D6- and D5-brane models is that
of the relation between anomaly cancellation and global 
and local tadpole cancellation. In these models non-Abelian
gauge anomalies are cancelled by the requirement of global tadpole
cancellation. $U(1)_\alpha G_\beta G_\beta$ mixed anomalies are
cancelled by a Green Schwarz mechanism together with tadpole
cancellation requirement. Finally, cubic anomalies are cancelled by
the Green Schwarz mechanism. One can draw the conclusion that tadpole 
cancellation is crucial for the overall consistency of these models. 
For a study of local anomalies in the context 
of type II B string constructions and their implications 
see reference \cite{Scrucca:2002is}.

Apart from ensuring anomaly cancellation, global tadpole  cancellation 
imposes an additional condition such that for a $U(N)$ group the number
of fundamental representations be equal to that of anti-fundamental
representations. This requirement applies even to $U(2)$ and $U(1)$ 
gauge groups with the consequence of restricting  the way we assign the 
quarks and leptons as $U(2)$ doubles and/or anti-doublets. If all quarks
were doublets and leptons anti-doublets, then the above condition
would not be respected. The only possibility to equal the 
number of doublets and anti-doublets is to have two families of
quarks as $U(2)$ doublets, with the third one and with the leptons as
$U(2)$ anti-doublets. This only works for {\it three} generations and
thus elegantly relates the number of colours to that of generations
\cite{imr}.

For the D6-brane models of \cite{imr} that we address in the
following, the aforementioned  couplings 
$c_i^\alpha B_i \wedge F_\alpha$  bring mass terms for three of the
initial four $U(1)_\alpha$. However 
\cite{imr} one linear combination of these $U(1)$  remains massless,
because its coupling to any RR two-form field $B_i$ vanishes,
independent of the parameters of the model. An additional appropriate 
condition imposed on these allows one to identify the generator of 
this massless $U(1)$ with that of hypercharge, which  remains a 
gauge symmetry of the model. 
The couplings $\sum_{i, \alpha} c_i^\alpha B_i \wedge F_\alpha$ 
provide a mass term in the action of the structure
\begin{equation}\label{massmatrix}
(M^2)_{\alpha\beta}=g_\alpha g_\beta M_S^2 
\sum_{i=1}^{3} c_i^\alpha c_i^\beta, \quad c_i^\alpha=
N_\alpha n_{\alpha j} n_{\alpha k} m_{\alpha i},\quad i\not=j\not=k\not=i,
\quad \alpha,\beta=a,b,c,d.
\end{equation}
where $g_\alpha$ is the coupling of associated $U(1)_\alpha$,
 $c_i^\alpha$ result from integrating for D6 branes
over 3-cycle $C_i\wedge F_\alpha$ with  $C_i$ RR 5-form, and $M_S$ is 
the string scale defined up to an overall  (volume dependent)
factor. The eigenvalues of this mass matrix (squared
masses) are positive \cite{Ghilencea:2002da} irrespective of the 
parameters of the models. The masses (relative to the string scale)
and associated eigenvectors   of the three massive $U(1)$ fields 
can be computed explicitly \cite{Ghilencea:2002da}.   Generic 
results for the masses are  within a factor of 10 above and 
below the string scale respectively, with the third mass of the order of $M_S$.
A fourth $U(1)$ field remains massless, and this is  identifiable with the
$U(1)_Y$ of hypercharge after an additional (model building)
constraint on the wrapping numbers.

The next step is that of including the effects of the electroweak 
symmetry breaking. For this we considered the minimal Higgs content 
predicted in this class of models \cite{Ghilencea:2002da} which is at 
least two Higgs doublets\footnote{This is a somewhat generic feature in 
string models \cite{Ghilencea:2002jd}.}. Such effects bring
small corrections to the massive  $U(1)$ fields
computed in the absence of EW symmetry breaking, since they are  
suppressed by
$\eta\equiv <\!v\!>^2\!\!\!/M_S^2$. There is however a more important aspect 
to be addressed. The mass of the $Z$ boson itself receives corrections
from mixing with the additional $U(1)$ fields, since the Higgs 
sector is also
charged under $U(1)_{b,c}$. Since the masses of the  $U(1)$ fields
are essentially of string origin, induced by their initial couplings
to $B_i$ fields, the mass of $Z$ boson will itself receive such a
correction in addition to its SM value. The correction 
to the mass of the $Z$ boson can be computed as an expansion 
in $\eta$ (see details in \cite{Ghilencea:2002da})
\begin{equation}\label{correction1}
M_Z^2=M_0^2 \left[1+\eta \xi_{21}+\eta^2 \xi_{31}+\cdots\right]
\end{equation}
with $M_0$ the SM $Z$ boson mass and $\xi_{21}$ given by (for $\xi_{31}$ 
see Appendix of \cite{Ghilencea:2002da}) 
\begin{eqnarray}\label{correction2}
\xi_{21} & = & -\left\{
\beta_1^2 \left[2 \beta_1 g_y^2 \,\nu\, n_{c1} (1+\cR^2)-(36 g_a^2+g_y^2
\cR^2)\beta_2 n_{a2} \nu\right]^2+
4 \beta_1^2 \beta_2^4 \epsilon^2 (36 g_a^2+g_y^2 \cR^2)^2\right.\nonumber\\
&&\!\!\!\!\!\!\!\!\!\!\!\!\!\!\!\!\!
+\left. 9\beta_2^4\left[ g_y^2 n_{b1} \,\nu\, \cR^2 +12 g_a^2 
[ 3 n_{b1}\,\nu\,-n_{c1}(1+\cR^2)\cos(2\theta)]\right]^2
\right\} 
\left[5184\beta_1^2 \beta_2^2 \epsilon^2 g_a^4 n_{c1}^2
(1+\cR^2)^2\right]^{-1}
\end{eqnarray}
where $\cR=g_d/g_c$ and $n_{a2}$ are model parameters, $n_{c1}$ and 
$n_{b1}$ are either 0 or $\pm 1$ function of the model, $\nu=1,1/3$, 
 $\beta_{1,2}=1,1/2$,  $g_a^2=g_{QCD}^2/6$ and $g_y$ is the
hypercharge coupling. $\theta$ is a mixing angle in
the Higgs sector (which does not affect significantly the predictions).
Correction (\ref{correction1})
must be compatible with electroweak scale 
measurements, and this may be ensured using $\rho$-parameter
constraints. From this a lower bound on the allowed values of the
string scale $M_S$ emerges. 
This bound  should not be too large compared to the  TeV scale
because the models we address are non-supersymmetric and one would otherwise
re-introduce a hierarchy problem. Detailed calculations show that
generic lower bounds on $M_S$ are within the range $1.5 - 5$ TeV  
\cite{Ghilencea:2002da} and are to a great extent parameter-independent.

The analysis so far has ignored the fact that the coefficients
$c_i^\alpha$ in eq.(\ref{massmatrix}) depend on the normalisation of
the kinetic terms of RR fields $B_i$. In  a canonical normalisation  
of the latter, extra volume factors $\xi_i$ appear in the definition of
$c_i^\alpha$. If equal for the three torii $\xi_i=\xi_j$,
$i,j=1,2,3$ (this can be easily respected if the ratio of the radii is
the same for all torii)  the volume factors can be 
``absorbed'' in the re-definition of the string scale $M_S$, as we
already did in eq.(\ref{massmatrix}). Therefore the lower  bounds on 
$M_S$ that we found in the region $1.5-5$ TeV  are subject to such 
additional corrections, difficult to evaluate numerically on string 
theory grounds.

One can take this analysis a step further \cite{Ghilencea:2002by}
by using the bounds on (the re-scalled) $M_S$ compatible with $\rho$ parameter 
constraints to obtain lower bounds on $U(1)$ masses. These will be independent 
of the volume factors mentioned    and represent a {\it
prediction} in this class of models.  Of particular importance
and experimental relevance is that of mixing of $Z$ eigenvector with 
any of the massive $U(1)$ fields, induced after electroweak symmetry 
breaking. This is usually  of order $10^{-3}$ \cite{Leike:1998wr} or 
possibly  less  \cite{Erler:1999nx}. Therefore, computing the eigenvectors
corresponding to the mass eigenstates after electroweak symmetry 
breaking is necessary  \cite{Ghilencea:2002by}, which also helps 
identify specific signatures of the $Z'$ bosons and compare  
them with other models.

Our results are presented in Table \ref{tablespectrum} and 
show that $Z'$ bosons in the TeV-region may be present in these models.
The bounds are somewhat larger than those of  their counterparts 
in (heterotic) string models or extensions of SM (MSSM) which are
in the range \cite{pdg} of: 
$Z'_{SM}$ boson defined to have the same 
couplings to fermions as the SM Z boson: $809$ GeV,
$Z_{LR}$ of $SU(3)_c\times
SU(2)_L\times SU(2)_R\times U(1)$: 564 GeV; 
$Z_\chi$ of $SO(10)\rightarrow
SU(5)\times U(1)_\chi$: 545 GeV;
$Z_\psi$ of $E_6\rightarrow SO(10)\times U(1)_\psi$: 
146 GeV.

\begin{table}[ht] 
\begin{center}
\begin{tabular}{|c|c|c|c|c|c|c|c|c|}
\hline
Higgs        & $\nu$ & $\beta_1$ & $\beta_2$ & $n_{c1}$ & $n_{b1}$ & $\cM_2\, (TeV)$  & $\cM_3\, (TeV)$  & $\cM_4\,(TeV)$  \\
\hline\hline  
$n_H=1,n_h=0$   & $1/3$   &  1/2  &   $1/2$  & $1$     &  $-1$   & $>25$  & $1.2 \, (0.75)$ &  $>3.5$  \\
$n_H=1,n_h=0$   & $1/3$   &  1/2  &   $1$    & $1$     &  $-1$   & $>110$ & $1.2\,(0.65)$   &  $>3.5$  \\
$n_H=1,n_h=0$   & $1/3$   &  1/2  &   $1/2$  & $-1$    &  $1$    & $>25$  & $1.7\,(0.75)$   &  $>3.5$  \\
$n_H=1,n_h=0$   & $1/3$   &  1/2  &   $1$    & $-1$    &  $1$    & $>100$ & $1.2\,(0.65)$   &  $>3.5$  \\
$n_H=0,n_h=1$   & $1/3$   &  1/2  &   $1/2$  & $1$     &  $1$    & $>25$  & $1.2\,(0.6)$    &  $>3.5$  \\
$n_H=0,n_h=1$   & $1/3$   &  1/2  &   $1$    & $1$     &  $1$    & $>100$ & $1.2\,(0.6)$    &  $>3.5$  \\
$n_H=0,n_h=1$   & $1/3$   &  1/2  &   $1/2$  & $-1$    &  $-1$   & $>25$  & $1.1\,(0.65)$    &  $>3.5$  \\
$n_H=0,n_h=1$   & $1/3$   &  1/2  &   $1$    & $-1$    &  $-1$   & $>65$  & $1.2\,(0.6)$    &  $>3.25$  \\

\hline\hline  
$n_H=1,n_h=1$   & $1$      &  1  &   $1/2$   & $1$    &  $0$   & $>1.5$   & $>1.5$ & $1\,(0.6)$     \\
$n_H=1,n_h=1$   & $1$      &  1  &   $1$     &  $1$   &  $0$   & $>4.5$   & $>1.5$ & $1.2\,(0.6)$   \\
$n_H=1,n_h=1$   & $1$      &  1  &   $1/2$   &  $-1$  &  $0$   & $>3$     & $>1.7$ & $0.75\,(0.6)$   \\
$n_H=1,n_h=1$   & $1$      &  1  &   $1$     &  $-1$  &  $0$   &  $>5.4$  & $>1.6$ & $1.1\,(0.55)$    \\
$n_H=1,n_h=1$   & $1/3$    &  1  &   $1/2$   &  $1$   &  $0$   &  $>3.5$  & $>1.6$ & $1.2\,(0.6)$    \\
$n_H=1,n_h=1$   & $1/3$    &  1  &   $1$     &  $1$   &  $0$   &  $>12.5$ & $>1.6$ & $1.125\,(0.6)$   \\
$n_H=1,n_h=1$   & $1/3$    &  1  &   $1/2$   &  $-1$  &  $0$   &  $>3.6$  & $>1.6$ & $1.1\,(0.575)$   \\
$n_H=1,n_h=1$   & $1/3$    &  1  &   $1$     &  $-1$  &  $0$   &  $>13$   & $>1.7$ & $1.125\,(0.6)$   \\
\hline 
\end{tabular}
\end{center}
\caption{\small Lower bounds on $U(1)$ masses $\cM_{1,2,3}$ for
minimal Higgs content:  one (top table) and two (lower table) 
pairs of Higgs doublets. The values correspond to any of Z bosons' 
mixing with massive U(1)'s  less than $1.5\times 10^{-3}$. Values 
in brackets correspond to mixing of up to $3\times 10^{-3}$.
These bounds apply for $g_d/g_c=\cO(1)$, 
otherwise they may increase further.}
\label{tablespectrum}
\end{table}

So far we only outlined part of the phenomenological 
implications of the D6 brane models. For a full picture of the 
viability of the models
there are additional issues to be addressed.  These include the
values of the gauge couplings at the TeV scale and the extent to which
these can be compatible with the electroweak scale measurements. 
One potential drawback of the 
models discussed here is that unification of the gauge couplings is
not possible (given the non-supersymmetric spectrum).
However this is a more generic problem, manifest in all models with a low
string scale, whether supersymmetric or not \cite{Ghilencea:2000dg}.
In the absence of unification, one should then ensure that the ratio of 
the couplings at the string scale is that obtained from the RG flow 
using low energy values as input. This is relevant since 
the value of gauge couplings at the string scale depends ultimately
on the volume each $D6$ brane is wrapping,  bringing additional 
constraints on the models.  There is one final complication because 
threshold effects induced by additional momentum   states 
can affect significantly  the value of the  couplings 
 \cite{Ghilencea:2002ff}, making the whole analysis rather 
difficult.

The class of models considered here for the
phenomenological analysis is free of tachyons and RR tadpoles,
but the models are non-supersymmetric and there will in general be NSNS 
tadpoles, which is generic in these  constructions. 
For the D6-brane models addressed one may also ask 
how one could achieve a low value for the string scale.
Since the models are non-supersymmetric, theory should fix 
this value in the few TeV region. (We have seen from the 
numerical analysis that low energy constraints as $\rho$ parameter
can be accommodated with a TeV string scale, which is 
welcome for the viability and  consistency of the models).
However, for the D6 brane models yielding the SM fermionic spectrum 
there is no dimension transverse to all  the
intersecting branes, to allow one to predict the right value
of 4D Planck scale.  The problem is not present in the D5 brane models 
mentioned in this talk, which yield  similar results 
for the  $U(1)$ masses and bounds on $M_S$ from the $\rho$
parameter \cite{Ghilencea:2002da}. One solution  for the
D6-brane models to have a  low string scale was suggested in
\cite{Angel.Uranga}. The idea is to construct intersecting 
D6 branes wrapped on 3-cycles localised in a small region of a
Calabi-Yau manifold. By enlarging the dimensions of the
Calabi-Yau which are transverse to the set of D6 branes the 4D Planck
mass is recovered even for low (``TeV-scale'') values of $M_S$. 
This would not require enlarging  the size of (local) 3-cycles 
(and would not bring about unobserved light Kaluza-Klein modes that 
would appear in such case).

To conclude, the class of D6 and D5 brane models addressed provide 
an interesting alternative to previous heterotic constructions
and  are successful in deriving the symmetry and the  fermionic
spectrum of the SM. They may thus  provide us an understanding of 
the string embedding of the Standard Model.  While such models are 
intensively studied 
at the string level, their phenomenological viability was 
not investigated systematically. Our attempt to address it
in \cite{Ghilencea:2002da},  \cite{Ghilencea:2002by}
(see also recent \cite{dcm}) shows that for a suitable choice of 
model parameters, the models may be able to comply with the low 
energy constraints considered there.

\vspace{0.6cm}
\noindent
{\bf Acknowledgements:} 
The author thanks  L.~E.~Ib\'a\~nez, F. Quevedo and N. Irges for their
collaboration. Helpful discussions on related topics  with S. F\"orste, 
G. Honecker, R.~Rabad\'an and A. Uranga are acknowledged.
This work was supported by  PPARC (U.K.).

\end{document}